\documentclass{caosp}

\usepackage{graphicx}

\articleNo{A03}
\pubyear{2014}
\volume{43}
\volnumber{3}
\firstpage{1}
\received{November 10, 2013}
\accepted{January 14, 2014}

\begin{document}

\htitle{Affordable spectroscopy for 1m-class telescopes\ldots}
\hauthor{B.\,Cs\'{a}k, J.\,Kov\'{a}cs, Gy.M.\,Szab\'{o}, L.L.\,Kiss, \'{A}.\,D\'{o}zsa, \'{A}.\,S\'{o}dor and I.\,Jankovics}

\title{Affordable spectroscopy for 1m-class telescopes: recent developments and applications}

\author{
        B.\,Cs\'{a}k\inst{1,2}
      \and
        J.\,Kov\'{a}cs\inst{1,2}
      \and
        Gy.M.\,Szab\'{o}\inst{1,2,3}
      \and
        L.L.\,Kiss\inst{2,3}
      \and
        \'{A}.\,D\'{o}zsa\inst{1,2}
      \and
        \'{A}.\,S\'{o}dor\inst{3}
      \and
        I.\,Jankovics\inst{1,2}
       }

\institute{
           Gothard Astrophysical Observatory and Multidisciplinary Research Center of Lor\'and E\"otv\"os University
         \and
           ELTE Gothard-Lend\"ulet Research Group\\9700 Szombathely, Szent Imre herceg u. 112., Hungary
         \and
           Konkoly Observatory, Research Centre for Astronomy and Earth Sciences, Hungarian Academy of Sciences\\1121 Budapest, Konkoly Th. M. \'ut 15-17., Hungary
          }

\date{November 10, 2013}

\maketitle

\begin{abstract}
Doppler observations of exoplanet systems have been a very expensive technique, mainly due to the high costs of high-resolution stable spectrographs. Recent advances in instrumentation enable affordable Doppler planet detections with surprisingly small optical telescopes.

We investigate the possibility of measuring Doppler reflex motion of planet hosting stars with small-aperture telescopes that have traditionally been neglected for this kind of studies. After thoroughly testing the recently developed and commercially available Shelyak eShel echelle spectrograph, we demonstrated that it is routinely possible to achieve velocity precision at the $100~\mathrm{m\,s}^{-1}$ level, reaching down to $\pm 50~\mathrm{m\,s}^{-1}$ for the best cases. We describe our off-the-shelf instrumentation, including a new 0.5m RC telescope at the Gothard Astrophysical Observatory of Lor\'and E\"otv\"os University equipped with an intermediate resolution fiber-fed echelle spectrograph.

We present some follow-up radial velocity measurements of planet hosting stars and point out that updating the orbital solution of Doppler-planets is a very important task that can be fulfilled with sub-meter sized optical telescopes without requesting very expensive telescope times on 2--4~m (or larger) class telescopes.

\keywords{stars: planetary systems -- techniques: radial velocities}
\end{abstract}

\section{Introduction}

Measuring high-precision Doppler velocities has been one of the most successful methods for finding substellar companions of main-sequence stars. The key ingredient in detecting the tiny reflex motion of planet hosting stars has been the stability of the spectrographs, coupled with novel techniques of wavelength calibration, such as using an iodine cell as wavelength reference (Marcy \& Butler, 1992). The mandated technique is therefore usually expensive (in terms of telescope time), that lead to two major limits in the allocation of observing time:
(1) Spectroscopic confirmation of newly discovered exoplanets faces increasing difficulties to get telescope time, especially for $<10~\mathrm{m\,s}^{-1}$ observations.
(2) Very limited time is allocated to revisiting of known exoplanet systems (both transiting and RV planets) to refine the orbital ephemeris and search for long-period companions.

In this paper, we show some of the first results of a novel approach to these issues. We equipped sub-meter category telescopes with a very cheap, off-the-shelf spectrograph ($R=11\,000$ resolving power), and got to the $\pm 50~\mathrm{m\,s}^{-1}$ velocimetric accuracy for stars brighter than $10^\mathrm{m}$. The resolution is also favorable to test correlations between the radial velocities and line profiles, e.g. bisector span. This accuracy will be enough to
(1) confirm the planetary nature of massive hot Jupiters;
(2) uncover false transiting candidates, e.g. systems with double stars, M and L dwarfs etc.;
(3) search for long-period modulations and discover distant massive companions in the systems.

\section{Observations and data analysis}

\subsection{Telescopes and spectrograph}

Observations were carried out between September 2011 and April 2013 at two locations in Hungary: at the Gothard Astrophysical Observatory of E\"otv\"os University, Szombathely (hereafter GAO) and at the Piszk\'estet\H{o} Mountain Station of Hungarian Academy of Sciences (hereafter PO). GAO has a newly installed 0.5m diameter $f/9$ RC telescope for spectroscopic observations. It has a MoFoD MkII fork mount built by Gemini Telescope Design. The telescope axes have a friction drive system with encoder feedback, resulting periodic and guiding error less than 1{\arcsec} during observations. At PO we have used the 1m RCC telescope.

The spectrograph was the same fiber-fed instrument at both locations, the eShel system of the French Shelyak Instruments (Thizy \& Cochard, 2011). The eShel system consists of three main components. The fiber injection and guiding unit is attached to the RC focus of the telescope. The field of view is monitored by a Watec 120N+ video camera through a tilted mirror. There is a $75~\mu\mathrm{m}$ diameter hole in the middle of the mirror, where the image of the observed star has to be placed and held during the observation. The fiber injector module is located behind the hole and injects the light into the $50~\mu\mathrm{m}$ core diameter multi-mode acquisition fiber.

The object fiber is connected to the echelle spectrograph unit in a thermally isolated room. The light from the object fiber is collimated by a 125~mm $f/5$ doublet achromat and projected to an R2 echelle grating. The grating is used between orders \#29 and \#56 typically. Cross-dispersion is made by a prism, then the echelle spectrum is imaged by a 85~mm focal length $f/1.8$ telephoto lens, and recorded by a QSI 532ws CCD camera (Kodak KAF1600 CCD chip).

The third, separated unit includes the ThAr calibration lamp and the flat-fielding light source, which is a combination of a Tungsten lamp and LED sources to cover the whole optical range with smooth and flat continuum light. The light of the calibration lamps is led to the fiber injection and guiding unit at the telescope through another, $200~\mu\mathrm{m}$ core diameter multi-mode fiber. During calibration exposures an electromagnetically controlled flip-mirror closes the light path from the telescope and projects the calibration light into the object fiber injection unit. The light sources and the flip-mirror are controlled from the observer computer via serial port interface. The image acquisition and the calibration unit is controlled with the freeware Audela\footnote{\tt http://www.audela.org} software.

\begin{figure}[htp]
\centerline{
\includegraphics[width=\textwidth]{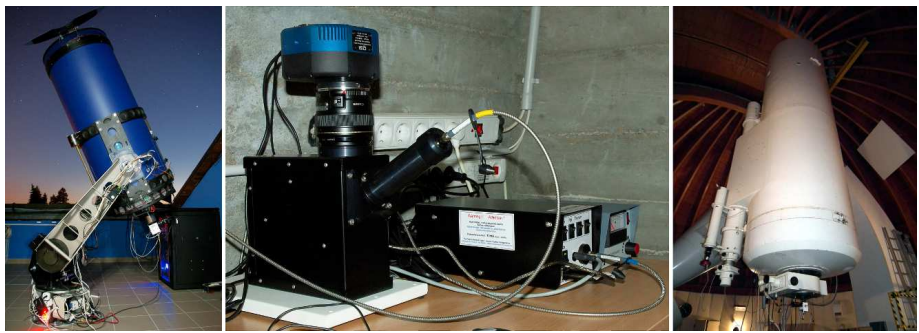}
}
\caption{{\it Left}: The 0.5m RC telescope of GAO used exclusively for spectroscopic observations. {\it Center}: The eShel echelle spectrograph and its accessories: the QSI~532ws CCD camera and the ThAr unit with its power supply box in a thermally isolated concrete room below the 0.5m RC telescope at GAO. {\it Right}: 1m RCC telescope of PO.}
\label{fig:instrumentation}
\end{figure}

The whole system is surprisingly compact and light and it is easy to carry and install on different telescopes (see Figure~\ref{fig:instrumentation}). Our copy of the fiber injector unit is optimized for a focal ratio $f/9$, so a Meade Series 4000 0.63x focal reducer was used at PO on the 1m $f/13.5$ RCC telescope.

\vspace*{-3.0ex}

\subsection{Data reduction and analysis}

Our data reduction pipeline is based on standard IRAF tasks and follows the standard steps of echelle reduction procedure implemented in IRAF. It is almost fully automated, only the identification of ThAr lines has to be done manually. The final result of the flow is a series of continuum normalized one dimensional spectra taken at the given night. The most crucial point of the reduction is the normalization and merging of the echelle orders because of the relatively high noise at their ends caused by blazing. This step is carried out according to the method given by W. Aoki\footnote{\tt http://optik2.mtk.nao.ac.jp/\~{}waoki/QL/specana200601e.pdf}.

\subsubsection{Radial velocity measurements}

For radial velocity measurements we use the IRAF {\tt rv.fxcor} task. Our spectra are cross correlated by templates from the synthetic library by Munari et al. (2005). The resolution of these templates is about $11\,500$, almost exactly the same as that of our spectra, therefore this library is ideal for us. Usually we chose the template that has nearly the same effective temperature and metallicity values as the object, but its $\log g$ and $v \sin i$ values are minimal. The CCR function is calculated between $4\,900$ and $6\,500$~\AA, excluding sodium doublet and telluric regions. The radial velocity and its error is estimated from a parabola fitting to the CCR function around its maximum. For checking the stability of our mount each night at least one radial velocity standard star is also observed.

\subsubsection{Signal to noise ratio}

\begin{figure}[htp]
\centerline{
\includegraphics[width=0.63\textwidth]{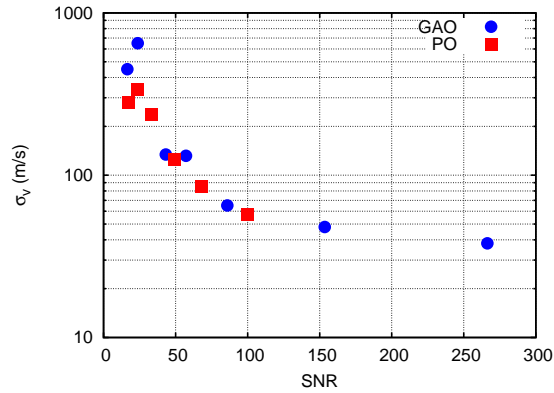}
}
\caption{Errors of the radial velocities plotted against the signal to noise ratios. The values were derived from the series of spectra with increasing exposure time of radial velocity standard star $\beta$ Vir. It can clearly be seen that SNR values higher than 150 give no further significant decrease in the RV error.}
\label{fig:betavir-snr-verr}
\end{figure}

Estimating the signal to noise ratio of our spectra is not a trivial task because of the many spectral orders and lines. The SNRs were measured in a relatively line-free spectral region between $6\,774$~\AA\ and $6\,780$~\AA\ as the root-mean-square of the deviations from the maximum flux. By good seeing conditions we can reach SNR of about 100 for an F9~V star of $V = 3.6^\mathrm{m}$ with 1 min exposure time using the spectrograph on our 0.5m RC telescope. The limiting magnitude is about $V = 10.5^\mathrm{m}$ with the 0.5m telescope and $V = 12^\mathrm{m}$ with the 1m telescope. To record spectra of such faint stars with still acceptable SNR, at least 1 hour of exposure time is needed. The accuracy of the derived RVs as a function of the SNR is shown in Figure~\ref{fig:betavir-snr-verr}.

\section{Some selected results}

Here we present our results on two HATNet exoplanets, namely HAT-P-2b and HAT-P-22b. The second planet is causing a radial velocity amplitude of about $300~\mathrm{m\,s}^{-1}$, so it is an ideal target for presenting the capabilities of our mount.

\subsection{HAT-P-2b}

The massive hot Jupiter HAT-P-2b was discovered by Bakos et al. (2007). Its size is around that of Jupiter (Bakos et al., 2007, P\'al et al., 2010), implying a density of $12~\mathrm{g\,cm}^{-3}$. The average density is consistent with an extended H-He atmosphere assuming a very massive core having $\approx 100$ Earth-mass. The system is unique due to the significant eccentricity of the planet, $e = 0.5171 \pm 0.0033$ (P\'al et al., 2010), therefore the transit light curve itself is asymmetric, that allows a photometric detection of the eccentricity. The insolation reaching the planet's surface varies by a factor of 9. Bakos et al. (2007) interpreted the large eccentricity as a likely hint for additional planets in the system. They also noted that RV curves taken at the Keck and Lick observatories are slightly inconsistent, and residuals of the RV curve are consistent with large jitter levels or perturbations from a third body.

\begin{figure}[htp]
\centerline{
\includegraphics[width=0.80\textwidth]{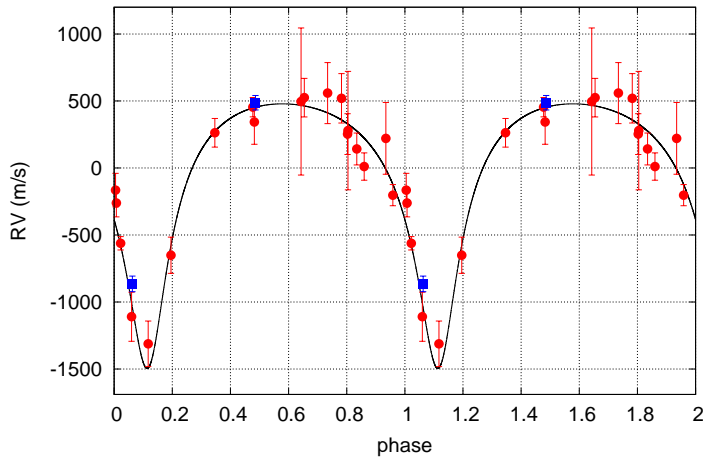}
}
\caption{A model RV curve of the exoplanet host star HAT-P-2 based on photometric observations overplotted with our (circles) and Mercator/HERMES (squares) radial velocity measurements. Note that the accuracy of our data is sometimes better than $100~\mathrm{m\,s}^{-1}$, which  clearly shows the applicability of our technique for detection exoplanets with a relatively small RV amplitude.}
\label{HAT-P-2}
\end{figure}

We have monitored HAT-P-2 during a three month long period in spring of 2012, both from GAO and PO, and recorded 69 spectra in 20 nights. Our RV values fit very well the model RV curve derived from the photometric observations with an accuracy sometimes clearly better than $100~\mathrm{m\,s}^{-1}$ (Figure~\ref{HAT-P-2}).

\subsection{HAT-P-22b}

HAT-P-22b was discovered by Bakos et al. (2011) around the brighter component of a probable CPM binary, exhibiting an M dwarf companion at 9$^{\prime\prime}$ separation. HAT-P-22b orbits a $V = 9.7^\mathrm{m}$ G5 dwarf with a period of $P = 3.2122200^\mathrm{d} \pm 0.000009^\mathrm{d}$. The host star has a mass of $0.92 \pm 0.03~M_{\sun}$, radius of $1.04 \pm 0.04~R_{\sun}$, $T_\mathrm{eff}=5302 \pm 80~\mathrm{K}$, and metallicity of $+0.24 \pm 0.08$. The planet has a mass of $2.147 \pm 0.061~M_\mathrm{J}$, and compact radius of $1.080 \pm 0.058~R_\mathrm{J}$. Besides two follow-up observations in 2009, no further light curves were published in the literature.

\begin{figure}[htp]
\centerline{
\includegraphics[width=0.80\textwidth]{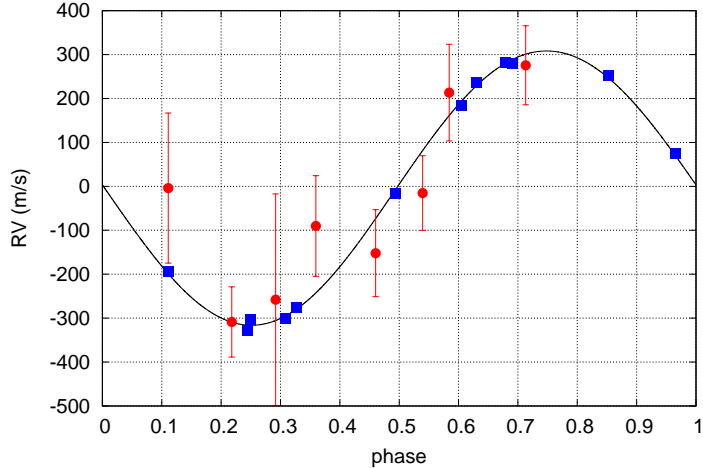}
}
\caption{Radial velocity data (squares) and the RV curve of the exoplanet host star HAT-P-22 based on Keck/HiRES spectra (Bakos et al., 2009). Our values are overplotted as circles. It can be seen that they fit well the measurements obtained from the observations taken by one of the world's largest telescopes.}
\label{HAT-P-22}
\end{figure}

Therefore 20 spectra of HAT-P-22 were taken by us during 6 nights also in spring of 2012. Despite the small amplitude caused by the planet, it could also have been detected just from our radial velocity measurements (Figure~\ref{HAT-P-22}). Based on Keck/HiRES and our data we have derived a new period of $P = 3.21164^\mathrm{d}$ and eccentricity of $e = 0.0136$ using the Systemic Console package (Meschiari et al., 2009; Meschiari \& Laughlin, 2010).

\section{Conclusions}

We can conclude that valuable spectroscopic results can be reached using off-the-shelf spectrographs attached to sub-meter telescopes. In good weather conditions and for bright stars the error of the radial velocities can be decreased down to $50~\mathrm{m\,s}^{-1}$ for a limited time, but for a longer period $100-200~\mathrm{m\,s}^{-1}$ is much more reliable. Such an accuracy is enough for spectroscopic follow-up measurements of stars hosting hot Jupiters, and very good for pulsating and binary/multiple stars. Our mount is also a very useful tool in teaching of spectroscopy, at both BSc and MSc levels.

\acknowledgements
This project has been supported by the Hungarian OTKA Grants K83790 and MB08C 81013, the "Lend\"ulet-2009" Young Researchers' Program and the J\'anos Bolyai Research Fellowship of the Hungarian Academy of Sciences, and the City of Szombathely under Agreement No.~S-11-1027.

\end{document}